\documentclass[PRD,reprint,twocolumn,showpacs,
amssymb, amsmath, aps, showpacs, nofootinbib, superscriptaddress]{revtex4}

\usepackage{multirow}
\usepackage{graphicx}
\usepackage{dcolumn}
\usepackage{bm}
\usepackage{threeparttable}
\usepackage{subfigure}
\usepackage{color,txfonts}
\usepackage{ulem}

\newcommand{\beq}{\begin{equation}}
\newcommand{\eeq}{\end{equation}}
\newcommand{\beqa}{\begin{eqnarray}}
\newcommand{\eeqa}{\end{eqnarray}}

\begin{document}

\title{Search for gamma-ray emission from eight dwarf spheroidal galaxy candidates discovered in Year Two of Dark Energy Survey with Fermi-LAT data}

\author{Shang Li}
\affiliation{Key Laboratory of Dark Matter and Space Astronomy, Purple Mountain Observatory, Chinese Academy of Sciences, Nanjing 210008, China}
\affiliation{University of Chinese Academy of Sciences, Yuquan Road 19, Beijing, 100049, China.}
\author{Yun-Feng Liang}
\affiliation{Key Laboratory of Dark Matter and Space Astronomy, Purple Mountain Observatory, Chinese Academy of Sciences, Nanjing 210008, China}
\affiliation{University of Chinese Academy of Sciences, Yuquan Road 19, Beijing, 100049, China.}
\author{Kai-Kai Duan}
\affiliation{Key Laboratory of Dark Matter and Space Astronomy, Purple Mountain Observatory, Chinese Academy of Sciences, Nanjing 210008, China}
\affiliation{University of Chinese Academy of Sciences, Yuquan Road 19, Beijing, 100049, China.}
\author{Zhao-Qiang Shen}
\affiliation{Key Laboratory of Dark Matter and Space Astronomy, Purple Mountain Observatory, Chinese Academy of Sciences, Nanjing 210008, China}
\affiliation{University of Chinese Academy of Sciences, Yuquan Road 19, Beijing, 100049, China.}
\author{Xiaoyuan Huang$^\ast$}
\affiliation{Physik-Department T30d, Technische Universit\"at M\"unchen, James-Franck-Stra\ss{}e, D-85748 Garching, Germany}
\affiliation{Key Laboratory of Dark Matter and Space Astronomy, Purple Mountain Observatory, Chinese Academy of Sciences, Nanjing 210008, China}
\author{Xiang Li$^\ast$}
\affiliation{Key Laboratory of Dark Matter and Space Astronomy, Purple Mountain Observatory, Chinese Academy of Sciences, Nanjing 210008, China}
\affiliation{University of Chinese Academy of Sciences, Yuquan Road 19, Beijing, 100049, China.}
\author{Yi-Zhong Fan$^\ast$}
\affiliation{Key Laboratory of Dark Matter and Space Astronomy, Purple Mountain Observatory, Chinese Academy of Sciences, Nanjing 210008, China}
\author{Neng-Hui Liao}
\affiliation{Key Laboratory of Dark Matter and Space Astronomy, Purple Mountain Observatory, Chinese Academy of Sciences, Nanjing 210008, China}
\author{Lei Feng}
\affiliation{Key Laboratory of Dark Matter and Space Astronomy, Purple Mountain Observatory, Chinese Academy of Sciences, Nanjing 210008, China}
\author{Jin Chang}
\affiliation{Key Laboratory of Dark Matter and Space Astronomy, Purple Mountain Observatory, Chinese Academy of Sciences, Nanjing 210008, China}

\date{\today}

\begin{abstract}
Very recently the Dark Energy Survey (DES) Collaboration has released their second group of Dwarf spheroidal (dSph) galaxy candidates.
With the publicly-available Pass 8 data of Fermi-LAT we search for $\gamma-$ray emissions from the
directions of these eight newly discovered dSph galaxy candidates. No statistically significant $\gamma-$ray signal has been found in the combined analysis of these sources. With the empirically estimated J-factors of these sources, the constraint on the annihilation channel of $\chi\chi \rightarrow \tau^{+}\tau^{-}$ is comparable to the bound set by the joint analysis of fifteen previously known dSphs with kinematically constrained J-factors for the dark matter mass $m_\chi>250$ GeV. In the direction of Tucana III (DES J2356-5935), one of the nearest dSph galaxy candidates that is $\sim 25$ kpc away, there is a weak $\gamma-$ray signal and its peak test statistic (TS) value for the dark matter annihilation channel $\chi\chi\rightarrow \tau^{+}\tau^{-1}$ is $\approx 6.7$ at $m_\chi \sim 15$ GeV. The significance of the possible signal likely increases with time.
More data is highly needed to pin down the physical origin of such a GeV excess.
\end{abstract}


\maketitle
\section{Introduction}
The nature of dark matter particles is still unknown and among various speculated particles weakly interacting massive particles (WIMPs) are the most popular
candidates \cite{Jungman:1995df, Bertone:2004pz, Hooper:2007qk, Feng:2010gw}. WIMPs may annihilate or decay and then produce stable high-energy particle pairs such as electrons/positrons, protons/antiprotons, neutrinos/anti-neutrinos, $\gamma-$rays and so on. The main goal of the so-called indirect detection experiments is to identify cosmic rays or $\gamma-$rays with a dark matter origin \cite{Jungman:1995df, Bertone:2004pz, Hooper:2007qk, Feng:2010gw}. The charged cosmic rays are deflected by the magnetic fields and their energy spectra would also be (significantly) modified during their propagation. As a result, the dark-matter origin of some cosmic ray anomalies$-$for example, the well-known electron/positron excesses \cite{Chang:2008aa, Adriani:2008zr, Adriani:2011xv, FermiLAT:2011ab, Aguilar:2013qda, Aguilar:2014fea} $-$ is hard to reliably establish.
The morphology of prompt $\gamma-$rays from annihilation or decay, instead, directly traces the dark matter spatial distribution and is therefore possible to choose regions in the sky with high dark matter density to investigate the dark matter properties. The annihilation signal is expected to be the brightest in the Galactic center but the astrophysical backgrounds are very complicated there \cite{Bertone:2004pz, Hooper:2007qk}. That is why the dark matter annihilation origin of the GeV excess in the inner Galaxy \cite{Goodenough:2009gk, 2009arXiv0912.3828V,Hooper:2010mq, Hooper:2010im, Abazajian:2012pn, Gordon:2013vta, Hooper:2013rwa} has not been widely accepted yet though its significance has been claimed to be as high as $\sim 40 \sigma$ \cite{Daylan:2014rsa} and this excess is found to be robust across a variety of models for the diffuse galactic $\gamma-$ray emission \cite{Zhou:2014lva, Calore:2014xka, Huang:2015rlu, TheFermi-LAT:2015kwa}.
The dwarf Spheroidal (dSph) galaxies are widely believed to be favorable targets with high signal-to-noise ratio \cite{Lake:1990du, Baltz:2004bb, Strigari:2013iaa}, because on the one hand these objects are very nearby and on the other hand they are far away from complicated emission regions. Several searches for gamma-ray emissions from dwarf galaxies detected by Sloan Digital Sky Survey (SDSS), which covers the northern-hemisphere \cite{Simon:2007dq}, and earlier experiments \cite{York:2000gk, McConnachie:2012vd} have been performed using Fermi-LAT data, and none of them reported a significant detection
\cite{Ackermann:2011wa, GeringerSameth:2011iw, Tsai:2012cs, Mazziotta:2012ux, 2012PhRvD..86b3528C, Ackermann:2013yva,Ackermann:2015zua,2015PhRvD..91h3535G,2015arXiv151000389B}.
The ongoing Dark Energy Survey (DES) \cite{Abbott:2005bi,2016arXiv160100329D} is instead a southern-hemisphere optical survey and in early 2015 the DES Collaboration released their first group of dSph galaxy candidates \cite{Bechtol:2015cbp, Koposov:2015cua}. Shortly after that, another
dSph galaxy candidate (Triangulum II) was discovered
with the data from the Panoramic Survey Telescope
and Rapid Response System (Pan-STARRS) \cite{Laevens:2015una} and a few additional candidates were reported by other collaborations \cite{martin15_hydra, kim15_pegasus, kim15_horologium, laevens15_3dsph}.
Though a reliable J-factor is not available for most newly-discovered sources, the velocity dispersion measurements strongly suggest that some sources (e.g. Triangulum II, Horologium II) are indeed dark matter-dominated dSphs \cite{kim15_horologium,kirby15_tri2}.
The analysis of the publicly-available Fermi-LAT Pass 7
Reprocessed data found moderate evidence for $\gamma-$ray emission from Reticulum 2 \cite{Geringer-Sameth:2015lua, Hooper:2015ula}
and the signal was found to be consistent with the Galactic GeV excess reported in \cite{Hooper:2010mq, Hooper:2010im, Abazajian:2012pn, Gordon:2013vta, Hooper:2013rwa, Daylan:2014rsa, Zhou:2014lva, Calore:2014xka, Huang:2015rlu, TheFermi-LAT:2015kwa}. Interestingly, later, the J-factor of Reticulum 2 is found to be among the largest of Milky Way dSphs \cite{2015ApJ...808L..36B}. The analysis of the Fermi-LAT Pass 8 data in the direction of Reticulum 2, however, just found a $\gamma-$ray signal with a largest local significance of $\sim 2.4\sigma$ for any of the dark matter masses and annihilation channels \cite{Drlica-Wagner:2015xua}. Very recently the DES Collaboration has released their second group of new dSph galaxy candidates \cite{Drlica-Wagner:2015ufc}.
In this work we search for possible $\gamma-$ray emission from the directions of these very recently-discovered dSph galaxy candidates by the DES collaboration (hereafter we call them the DES Y2 dSph galaxy candidates).

\section{Data analysis}

In this paper, we used the newly released Pass 8 data to search for gamma-ray emission from these DES Y2 dSph galaxy candidates. The Pass 8 data benefit from an improved energy reach (changing from the range of $0.1-300$ GeV to $60~{\rm MeV}-500~{\rm GeV}$), effective area in particular in the low energy range, and the point-spread function \cite{Atwood:2013rka}. Thanks to such improvements, the differential point-source sensitivity improves by 30-50\% in P8R2\_SOURCE\_V6 data relative to P7REP\_SOURCE\_V15 data \cite{Drlica-Wagner:2015xua}, which make it more sensitive to faint sources like dSph galaxies. We used the Fermi-LAT data collected from 2008 August 4 to 2015 August 4 that have passed the P8R2 SOURCE event class selections from 500 MeV to 500 GeV.  To suppress the effect of the Earth's limb, the $\gamma-$ray events with zenith angles greater than $100^{\circ}$ were rejected. We use the updated standard Fermi Science Tools package with version v10r0p5 to analyze Fermi-LAT data. The regions of interest (ROI) are selected as regions centered at the position of each DES Y2 dSph galaxy candidate.
The selected data using criterion described above were divided into 100$\times$100 spatial bins with 0.1$^{\circ}$ bin size. Following Fermi team's recommendation, we adopted a diffuse emission model based on the Pass 7 Reprocessed model for Galactic diffuse emission but has been scaled to account for differences in energy dispersion between Pass 7 reprocessed data and Pass 8 data \footnote{http://fermi.gsfc.nasa.gov/ssc/data/access/lat/BackgroundModels.html}.

We took the approach developed in \cite{Tsai:2012cs, Ackermann:2013yva, Ackermann:2015zua} to analyze the gamma-ray emission for each dSph candidate, we refer readers to these literature for details of the approach we used.
First, we carried out a standard binned likelihood fit over the entire energy range with 24 logarithmically spaced energy bins to determine the background sources using {\it{gtlike}} tool in Fermi Science Tools.
All point-like sources from 3FGL \cite{Acero:2015hja} within $15^{\circ}$ of the center of each dSph galaxy were included and new excesses in TS map with ${\rm TS}\geq 25$ were identified as new sources and then included in the fit. The normalization of point sources within $5^{\circ}$ and the two diffuse backgrounds (Galactic diffuse emission and an isotropic components) were set free, while other parameters were fixed at the 3FGL values. No component associate with dSph was included in this step.

Next, we adopted a bin-by-bin analysis as \cite{Tsai:2012cs, Ackermann:2013yva, Ackermann:2015zua}.
For each ROI, a point source is added to the best fitting model in last step in the position of each dSph galaxy to consider the signal from the given direction.
We modeled these dSph galaxy candidates as point-like sources rather than spatially extended sources due to the lack of the information of the spatial extension of dark matter halos of these newly-discovered objects.
Likelihood profile, which is a curve of how likelihood varying as the flux of the newly added putative point source, are generated for everyone of 24 logarithmically evenly spaced energy bins.
Within each energy bin, we fixed all the model parameters but the normalization of the newly added dSph point source, and use a power-law spectral model ($dN/dE \propto E^{-\Gamma}$) with spectral index of $\Gamma=2$ to fit the putative dwarf galaxy source.
We scanned the likelihood as a function of the flux normalization of the assumed dark matter signal independently by varying the flux normalization to derive the profile.
These likelihood profile will be used in later analysis in section \ref{sec2_1}.
Using these profiles, we can also derive bin-by-bin energy-flux upper limits at 95\% confidence level for each dSph candidate, which are shown in Figure 1.

\begin{center}
\begin{table*}
\caption{DES Y2 dSph Candidates and the Estimated J-factors}
\begin{tabular}{cccc}
\hline
\hline
 Name &  $(l,b)^{\rm a}$ & ${\rm Distance^{b}}$ &  $\log_{10}{{\rm (Est. J)}^{\rm c}}$\\
     & (deg) &(kpc)& $\log10{\rm (GeV^{2}cm^{-5})}$\\
\hline
DES J2204-4626 &(351.15,-51.94) & $53\pm 5$ & 18.8\\
DES J2356-5935 &(315.38,-56.19) & $25\pm 2$ & 19.5\\
DES J0531-2801&(231.62,-28.88) & $182\pm 18$ & 17.8\\
DES J0002-6051&(313.29,-55.29) & $48\pm 4$ & 18.9\\
DES J0345-6026&(273.88,-45.65) & $92\pm 13$ & 18.3\\
DES J2337-6316&(316.31,-51.89) & $55\pm 9$ & 18.8\\
DES J2038-4609&(353.99,-37.40) & $214\pm 16$ & 17.6\\
DES J0117-1725&(156.48,-78.53) & $30\pm 3$ & 19.3\\
\hline
\end{tabular}
\begin{tablenotes}
\item $^{\rm a}$ Galactic longitudes and latitudes are adopted from \cite{Drlica-Wagner:2015ufc}.
\item $^{\rm b}$ The distances are taken from \cite{Drlica-Wagner:2015ufc}.
\item $^{\rm c}$ J-factors are estimated with the empirical relation $J(d)\approx 10^{18.3\pm 0.1}(d/100~{\rm kpc})^{-2}$ \cite{Drlica-Wagner:2015xua}.
\end{tablenotes}
\end{table*}
\end{center}

\begin{figure*}[t]
\begin{center}
\includegraphics[width=1.0\textwidth]{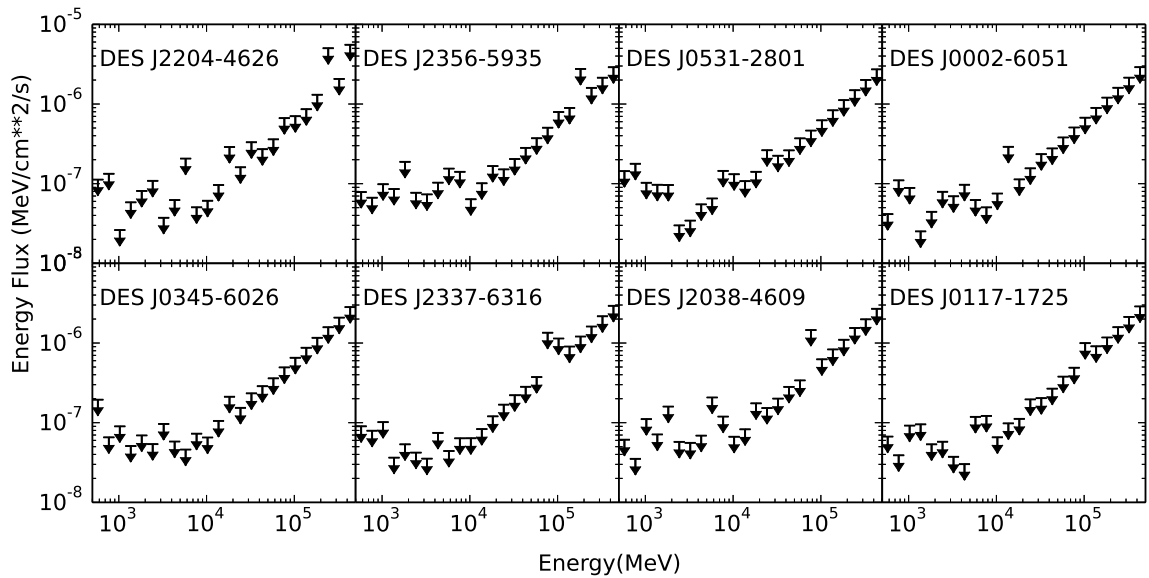}
\end{center}
\caption{Bin-by-bin integrated energy-flux upper limits at 95\% confidence level for each dwarf spheroidal satellite candidates that reported in \cite{Drlica-Wagner:2015ufc}.}
\label{Fig.1}
\end{figure*}

\subsection{Combined constraint on the dark matter physical properties with the newly discovered eight dwarf galaxies}
\label{sec2_1}
The dSph galaxies are known to be dominated by dark matter.
The ${\gamma}-$ray flux expected from annihilation of dark matter particles in a dSph galaxy is given by  \cite{Jungman:1995df, Bertone:2004pz, Hooper:2007qk, Feng:2010gw}
\begin{equation}
{\Phi}(E)={\frac{<{\sigma}v>}{8{\pi}m_{\chi}^{2}}\times \frac{dN_{\gamma}}{dE_{\gamma}}\times J},
\end{equation}
where ${m_{\chi}}$ is the rest mass of the dark matter particle, ${<{\sigma}v>}$ is the thermal average annihilation cross section, $dN_{\gamma}/dE_{\gamma}$ is the spectrum of prompt ${\gamma}-$rays resulting in dark matter particle annihilation and $J={\int}dld{\Omega}{\rho}(l)^{2}$ is the line-of-sight integral of the square of the dark matter density (i.e., the so-called J-factor).

\begin{figure*}[t]
\begin{center}
\includegraphics[width=0.45\textwidth]{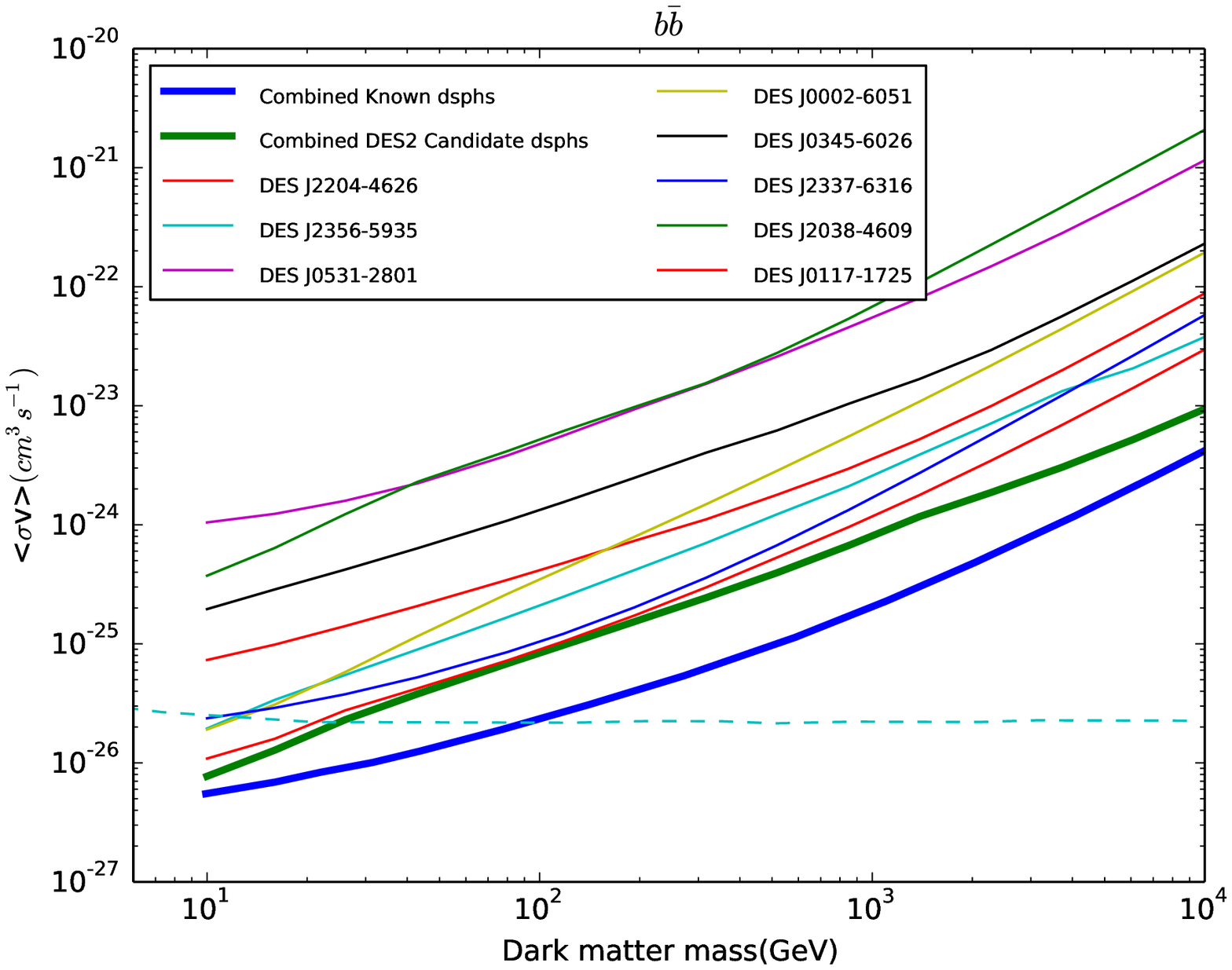}
\includegraphics[width=0.45\textwidth]{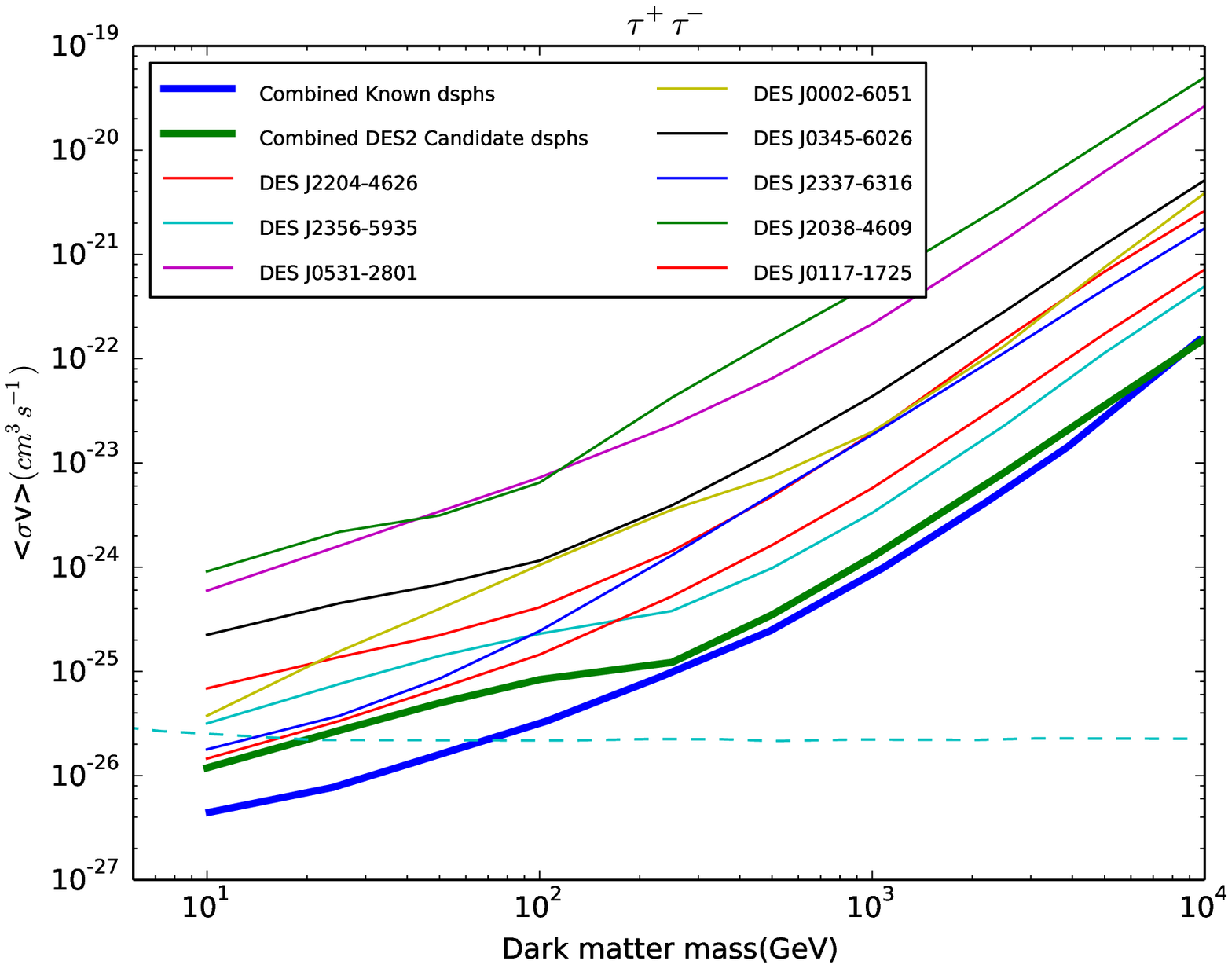}
\end{center}
\caption{The bounds on the $<{\sigma}v>$ at 95\% confidence level for dark matter annihilation to $b\bar{b}$ (left) and ${\tau^{+}\tau^{-}}$ (right) set by the individual DES Y2 candidate dSph as well as by the combined sample, separately.
The best limits inferred from a joint analysis of fifteen previously known dSphs with kinematically constrained J-factors \cite{Ackermann:2015zua}
 is plotted for comparison (see the thick blue line) and the thermal relic cross section \cite{Steigman:2012nb} is also shown for reference
 (see the dashed light blue curve).
}
\label{Fig.2}
\end{figure*}

\begin{figure*}[t]
\begin{center}
\includegraphics[width=0.45\textwidth]{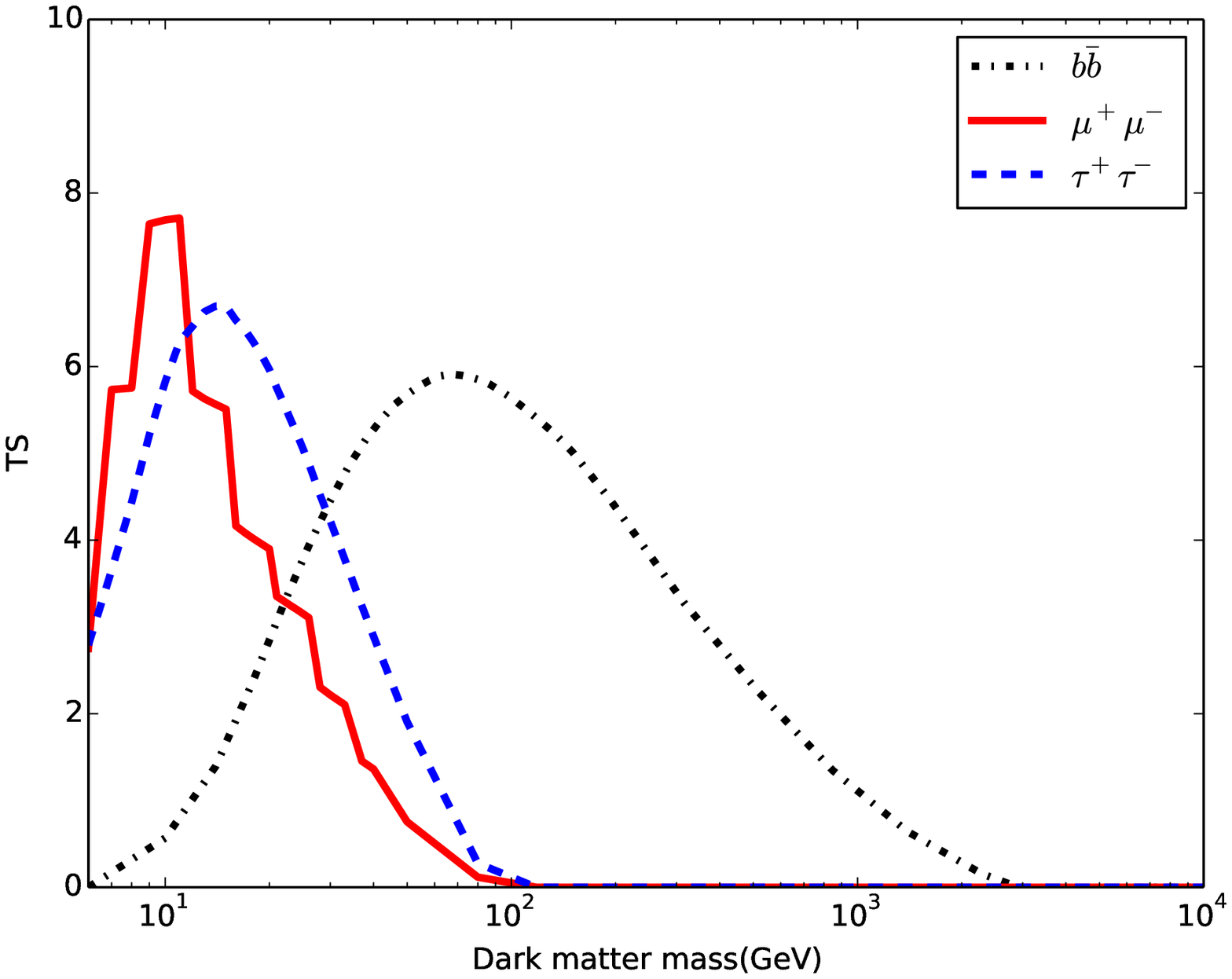}
\includegraphics[width=0.45\textwidth]{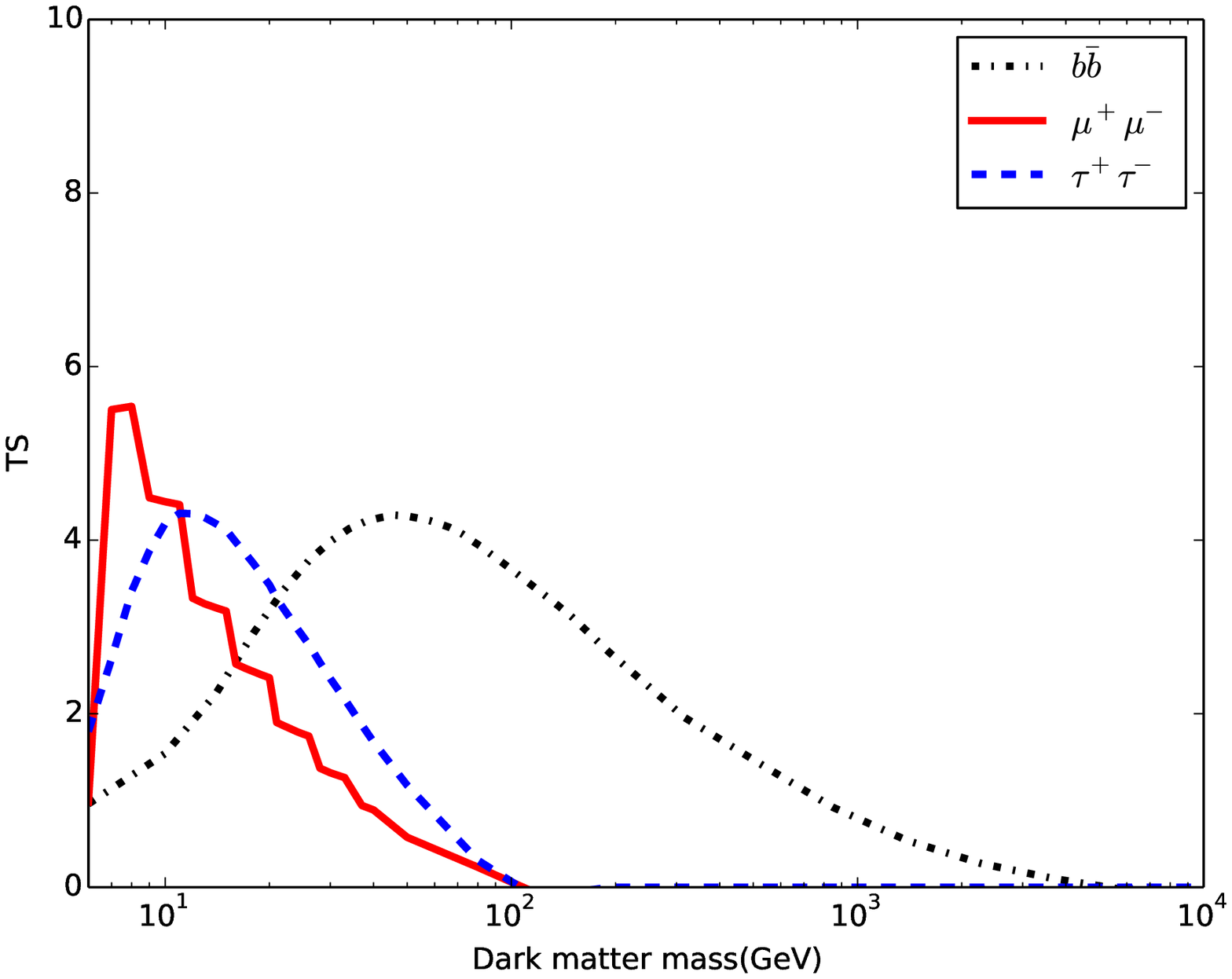}
\end{center}
\caption{The TS value of the $\gamma-$ray signal in the direction of Tucana III for different dark matter particle masses and annihilation channels in two different time intervals: The left panel is for 7 year Fermi-LAT data (2008 August 4 to 2014 August 4) and the right panel is for 3.5 year Fermi-LAT data (2008 August 4 to 2012 February 4).}
\label{Fig.3}
\end{figure*}

Utilizing the likelihood profile derived above, we reconstructed a broadband likelihood function by multiplying the bin-by-bin likelihood functions evaluated at the predicted fluxes for a given dark matter model.
Then we combined the eight DES Y2 dSph candidates' broad-band likelihood functions and added an extra J-factor likelihood term for each dSph candidate  to take into account the J-factor's statistical uncertainties.
The J-factor likelihood term for each dwarf galaxy is given by
\begin{equation}
L_{\rm J}(J_{\rm obs,i},{\sigma}_{\rm i})={1 \over \ln(10)J_{\rm obs,i}\sqrt{2\pi}{\sigma}_{\rm i}}
         \exp^{-[\log_{10}(J_{\rm i})-\log_{10}(J_{\rm obs,i})]^{2}/{2\sigma_{\rm i}^2}},
\end{equation}
where $i$ represent different target, $J_{\rm i}$ is the J-factor's ``real" value and the $J_{\rm obs,i}$ is the J-factor's empirically-estimated value with an error of ${\sigma}_{\rm i}$ \cite{Ackermann:2015zua}. After combining the J-factor likelihood term and the broad-band likelihood functions, the likelihood function for target $i$ reads
${\widetilde  L_{\rm i}}(\boldsymbol{\mu},\boldsymbol{\theta}_{\rm i}={\lbrace}\boldsymbol{\alpha}_{\rm i},J_{\rm i}{\rbrace}{\vert}D_{\rm i})=L_{\rm i}(\boldsymbol{\mu},\boldsymbol{\theta}_{\rm i}{\vert}D_{\rm i}){L_{\rm J}(J_{\rm obs,i},{\sigma}_{\rm i})}$,
where $\boldsymbol{\mu}$, $\boldsymbol{\alpha}_{\rm i}$, ${J_{\rm i}}$ and $D_{\rm i}$ represent the parameters of the dark matter model, the parameters of astrophysical background, the dSph J-factor and the gamma-ray data, respectively; and $\boldsymbol{\theta}_{\rm i}$ incorporates $\boldsymbol{\alpha}_{\rm i}$ and ${J_{\rm i}}$ \cite{Ackermann:2015zua}. To reduce the uncertainty on the direction of gamma-rays, we took into account four PSF event types (PSF0, PSF1, PSF2 and PSF3) when constructing the likelihood function, for which the broadband likelihood function for target $i$ is given by
$L_{\rm i}(\boldsymbol{\mu},\boldsymbol{\theta}_{\rm i}{\vert}D_{\rm i})={\prod\limits_{\rm j}}L_{\rm i}(\boldsymbol{\mu},{\boldsymbol\theta}_{\rm i}{\vert}D_{\rm i,j})$, where $j$ represents the different PSF event type \cite{Ackermann:2015zua}.

\begin{figure}
\begin{center}
\includegraphics[width=0.5\textwidth]{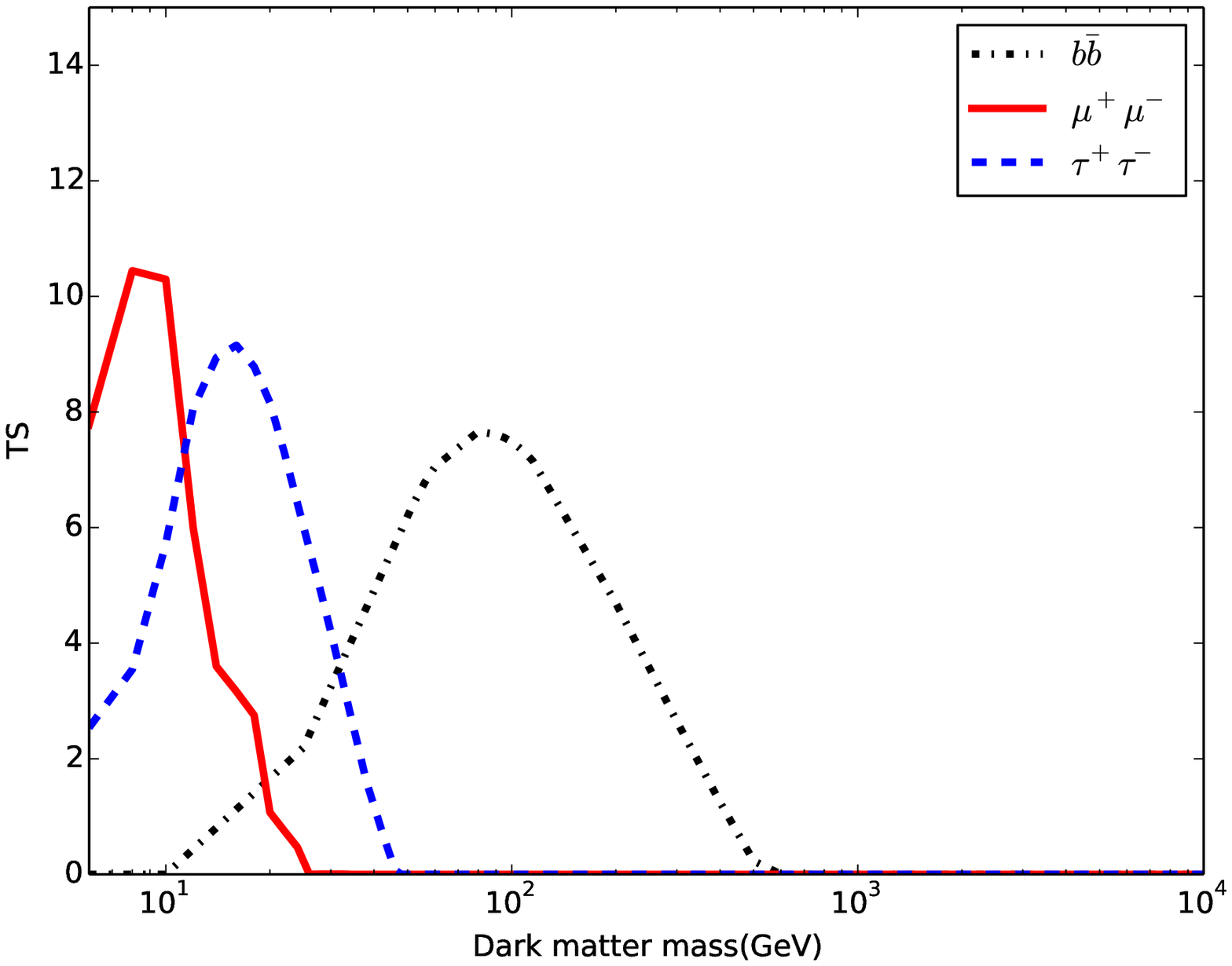}
\end{center}
\caption{The TS value of the possible dark matter annihilation signal in the combined $\gamma-$ray data in the directions of seven ``nearby" dSph galaxies (candidates), including Segue I, Segue II, Ursa Major II, Reticulum II, Tucana III, Cetus II and Willman I. The dark matter annihilation channels are labeled in the plot.}
\label{Fig.4}
\end{figure}

So far the reliable J-factors for these eight DES Y2 dSph galaxy candidates are unavailable. An empirical relation between the heliocentric distances
and J-factors of ultra-faint and classical dwarf galaxies is suggested to be $J(d)\approx 10^{18.3\pm 0.1}(d/100~{\rm kpc})^{-2}$ in \cite{Drlica-Wagner:2015xua}, where $d$ is the distance of the object to the Sun and a symmetric logarithmic uncertainty on
the J-factor of $\pm0.4$ dex for each DES dSph galaxy candidate is assumed \cite{Drlica-Wagner:2015xua}.
The estimated values of J-factors of the eight DES Y2 dSph galaxy candidates are presented in Table 1.
The individual and combined constraints on the dark matter annihilation channels of $\chi\chi \rightarrow b\bar{b}$ or $\tau^{+}\tau^{-}$ with these sources are presented in Figure 2. If the real J-factors are similar to our estimates, we can rule out the thermal relic cross section for WIMP with $m_\chi \lesssim 25$ GeV annihilating into either $b\bar{b}$  or  $\tau^{+}\tau^{-}$. We have also analyzed all the 16 dSph candidates reported in \cite{Bechtol:2015cbp, Koposov:2015cua, Drlica-Wagner:2015ufc} and found out that the combined constraints on the dark matter models are similar to that set by the DES Y2 data. Hence we do not present them in this work.

\subsection{$\gamma-$ray emission in the direction of Tucana III}
At a distance of 25 kpc, Tucana III (also known as DES J2356-5935) is one of the nearest dSph galaxy candidates locating at a high latitude that is suitable for dark matter indirect detection.  We first used a global binned likelihood fit in the  energy range from 500MeV to 500GeV  with a power-law spectral model (i.e., $\frac{dN}{dE} {\propto}E^{-2}$) for this dSph candidate.
Interestingly we found a weak ``excess" of gamma-ray in the direction of Tucana III with  ${\rm TS}\approx6.0$ after adding a possible weak point source (${\rm ra \approx 0.74^{\circ},~dec \approx -59.72^{\circ}}$) that is about 0.8 degree away.
In addition, we used bin-by-bin method to make a further analysis of Tucana III. In Figure 3 we present TS values of $\gamma-$ray signal in the direction of Tucana III for various annihilation channels and dark matter masses. Note that Fig.3a is for the 7 year LAT data (i.e., from 2008 August 4 to 2015 August 4), in which one can find that in each channel the significance of the signal peaks above $2\sigma$. Particularly, in the case of $\chi\chi\rightarrow \tau^{+}\tau^{-}$, the TS value of the fit peaks about 6.7  at $m_\chi \sim 15$ GeV. For the adopted empirical J-factor, a $\langle \sigma v\rangle_{\chi\chi\rightarrow \tau^{+}\tau^{-}}\sim 5\times 10^{-27}~{\rm cm^{3}~s^{-1}}$  is needed to reproduce the signal. In the case of $\chi\chi\rightarrow b\bar{b}$, the TS value of the fit peaks about 6 at $m_\chi \sim 66$ GeV. With the adopted empirical J-factor, a $\langle \sigma v\rangle_{\chi\chi\rightarrow b\bar{b}}\sim 2\times 10^{-26}~{\rm cm^{3}~s^{-1}}$  is needed to account for the signal. Such $m_\chi$ and $\langle \sigma v\rangle$ are similar to that ``preferred" by the Galactic GeV excess data as well as the possible gamma-ray signal in the direction of Reticulum 2 \cite{Hooper:2010mq, Hooper:2010im, Abazajian:2012pn, Gordon:2013vta, Hooper:2013rwa, Daylan:2014rsa, Zhou:2014lva, Calore:2014xka, Huang:2015rlu, TheFermi-LAT:2015kwa, Geringer-Sameth:2015lua, Hooper:2015ula}.  

In view of the similar though weak signals in the directions of Reticulum 2 \citep{Geringer-Sameth:2015lua, Hooper:2015ula, Drlica-Wagner:2015xua} and Tucana III, following the same data analysis approach we make a combined analysis for these two nearby dSph candidates. The TS values for a dark matter annihilation signal (for the representative channels $\chi\chi \rightarrow b\bar{b}$, $\tau^{+}\tau^{-}$ and $\mu^{+}\mu^{-}$) are evaluated. Interestingly, the TS values of this GeV-excess like signal increase sizeably and in the case of $\chi\chi\rightarrow \tau^{+}\tau^{-}$ we have the largest ${\rm TS}\approx 14$ for $m_{\chi} \approx 16$ GeV.
This corresponds to a local significance of $\sim3.7\sigma$, which decreases to $\sim2.3\sigma$ if we take into account the so-called trail-factor correction since here we have just chosen two sources from in total 16 DES dSph candidates.
We have also analyzed the gamma-ray emission in the directions of dSph galaxies (candidates) within a distance $\leq 40$ kpc from the sun, including Segue I, Segue II, Ursa Major II, Reticulum II, Tucana III, Cetus II and Willman I  but excluding Sagittarius and Canis Major since they are close to the Galactic plane. In the case of $\chi\chi\rightarrow \tau^{+}\tau^{-}$ we have the largest ${\rm TS}\approx 9.2$ at $m_{\chi} \approx 16$ GeV (see Figure 4).

Now we briefly examine the possible astrophysical origin of the weak $\gamma-$ray signal in the direction of Tucana III.
The ``signal" is too weak to directly get the variability information. Instead, we calculate the TS values of the potential `GeV excess' component in another time interval from 2008 August 4 to 2012 February 4 (i.e., the 3.5 year Fermi-LAT data) and the results are presented in Fig.3b. Interestingly the TS values of the annihilation channels shown in Fig.3a are larger than those in Fig.3b, implying that the significance is indeed increasing. Such an increase is expected in the models of dark matter annihilation or alternatively a steady astrophysical source.
It is well known that radio loud active galactic nuclei (RLAGNs) could be possible counterparts because of the high galactic latitude of the possible $\gamma-$ray signal. We note that there is a radio source PMN J2355-5948 about $\rm 0.3^{\circ}$  away from the optical position of Tucana III. It is included in the Parkes-MIT-NRAO (PMN) surveys \cite{1994ApJS...91..111W} and Sydney University Molonglo Sky Survey (SUMSS) \cite{Mauch:2003zh} and the fluxes at 4.85~GHz and 843~MHz are $55\pm8$ mJy and $259\pm8$ mJy, respectively. Assuming its radio emission follows a power-law distribution, the radio spectrum index can be estimated as $\alpha_{\rm r}\simeq 0.9$ (note that we refer to a spectral index $\alpha$ as the energy index such that $F_{\nu}\propto\nu^{-\alpha}$). Since blazars characterized by the flat radio spectrum ($|\alpha_{\rm r}|\leq$ 0.5) are dominated the extragalactic $\gamma$-ray sky and $\gamma-$ray emissions from only a handful of steep radio spectrum RLAGNs have been detected \cite{Ackermann:2015yfk, Liao:2015jfj}, it is less likely that PMN J2355-5948 is capable to produce significant $\gamma-$ray emission.

\section{Discussion and conclusion}
Dwarf spheroidal galaxies are one of the best targets for the indirect detection of dark matter annihilation signal. However, the reliable  identification of dwarf spheroidal galaxies in optical is a hard job. Before 2015, just 25 dSphs have been reported \citep{Simon:2007dq, York:2000gk, McConnachie:2012vd} and the $\gamma-$ray data analysis of these sources have imposed very stringent constraints on parameters for dark matter annihilation \citep{Ackermann:2011wa, GeringerSameth:2011iw, Tsai:2012cs, Mazziotta:2012ux, Ackermann:2013yva, Ackermann:2015zua}. In 2015, with the optical imaging data from Dark Energy Survey, 16 new dwarf spheroidal galaxy candidates, including a few ``nearby" sources at  distances of $20-30$  kpc, have been released \citep{Bechtol:2015cbp, Koposov:2015cua, Drlica-Wagner:2015ufc}. The sample of dSphs thus increased significantly and quickly. Although the reliable estimates of J-factors of most of these new dSph candidates are still unavailable, it is the time to carry out the $\gamma-$ray data analysis to check whether there are some interesting signals or not. The $\gamma-$ray search for the first group DES dSph candidates have been reported in \cite{Drlica-Wagner:2015xua}. No significant gamma-ray emission signal has been identified and strong constraints on the dark matter annihilation channels have been provided by adopting an empirical relation between the J-factor of the dSph and its distance to us \cite{Drlica-Wagner:2015xua}. A very weak signal resembling the Galactic GeV excess, however, may present in the direction of Reticulum 2 \cite{Geringer-Sameth:2015lua, Hooper:2015ula,Drlica-Wagner:2015xua}.

In this work we have analyzed the publicly-available Pass 8 data of Fermi-LAT in the directions of eight new dSph galaxy candidates discovered in Year Two of Dark Energy Survey (see Fig.\ref{Fig.1}). No statistically significant $\gamma-$ray signal has been found in the combined analysis of these new sources. With the empirically estimated J-factors of these sources, the constraint on the annihilation channel of $\chi\chi \rightarrow \tau^{+}\tau^{-}$ is found comparable to the bound set by the joint analysis of fifteen previously known dSphs with kinematically constrained J-factors for $m_\chi>250$ GeV (see Fig.\ref{Fig.2}). Interestingly, in the direction of Tucana III, a dSph galaxy candidates that is $\sim 25$ kpc away, there is a very weak GeV-excess like $\gamma-$ray signal. We have a ${\rm TS}\approx 6.7$ for the annihilation channel $\chi\chi\rightarrow \tau^{+}\tau^{-1}$  and $m_\chi \approx 15$ GeV. The significance of the possible signal increases with time (see Fig.\ref{Fig.3} for the comparison of the results for the 7 year and 3.5 year Fermi-LAT data), as expected in the models of dark matter annihilation or alternatively a steady astrophysical source. To further check the significance of the possible gamma-ray signal, we have also analyzed the Fermi-LAT Pass 8 data in the directions  of seven ``nearby" dSphs, including Segue I, Segue II, Ursa Major II, Reticulum II, Tucana III,  Cetus II and Willman I. In the case of $\chi\chi\rightarrow \tau^{+}\tau^{-}$ we have the largest ${\rm TS}\approx 9.2$ at $m_{\chi} \approx 16$ GeV for the combined $\gamma-$ray data set (see Fig.\ref{Fig.4}). Interestingly, the corresponding mass and annihilation cross section of dark matter for the weak gamma-ray signal
are consistent with those needed for the dark matter interpretation of GeV excess.
The origin of GeV excess from Galaxy inner region is still in heavy debate \citep{Cirelli:2015gux} and additional support to the dark matter interpretation could be from dSph galaxies that do not suffer from the contamination caused by the complicated background emission. Though our current results seems encouraging, we would like to remind that the astrophysical origin or even a statistical fluctuation origin of the very weak signal is still possible. More data is highly needed to draw a more formal conclusion.


\acknowledgments We thank the anonymous referee for helpful comments/suggestions. This work was supported in part by 973 Program of China under grant 2013CB837000, by NSFC under grants 11525313 (i.e., the National Natural Fund for Distinguished Young Scholars), 10925315, 11361140349 and 11103084, by Foundation for Distinguished Young Scholars of Jiangsu Province, China (No. BK2012047), and by the Strategic Priority Research Program (No. XDA04075500).

$^\ast$Corresponding authors (huangxiaoyuan@gmail.com, xiangli@pmo.ac.cn, yzfan@pmo.ac.cn).

\bibliographystyle{apsrev}
\bibliography{refs}

\begin{thebibliography}{60}
\expandafter\ifx\csname natexlab\endcsname\relax\def\natexlab#1{#1}\fi
\expandafter\ifx\csname bibnamefont\endcsname\relax
  \def\bibnamefont#1{#1}\fi
\expandafter\ifx\csname bibfnamefont\endcsname\relax
  \def\bibfnamefont#1{#1}\fi
\expandafter\ifx\csname citenamefont\endcsname\relax
  \def\citenamefont#1{#1}\fi
\expandafter\ifx\csname url\endcsname\relax
  \def\url#1{\texttt{#1}}\fi
\expandafter\ifx\csname urlprefix\endcsname\relax\def\urlprefix{URL }\fi
\providecommand{\bibinfo}[2]{#2}
\providecommand{\eprint}[2][]{\url{#2}}

\bibitem[{\citenamefont{Jungman et~al.}(1996)\citenamefont{Jungman,
  Kamionkowski, and Griest}}]{Jungman:1995df}
\bibinfo{author}{\bibfnamefont{G.}~\bibnamefont{Jungman}},
  \bibinfo{author}{\bibfnamefont{M.}~\bibnamefont{Kamionkowski}},
  \bibnamefont{and} \bibinfo{author}{\bibfnamefont{K.}~\bibnamefont{Griest}},
  \bibinfo{journal}{Phys. Rept.} \textbf{\bibinfo{volume}{267}},
  \bibinfo{pages}{195} (\bibinfo{year}{1996}), \eprint{hep-ph/9506380}.

\bibitem[{\citenamefont{Bertone et~al.}(2005)\citenamefont{Bertone, Hooper, and
  Silk}}]{Bertone:2004pz}
\bibinfo{author}{\bibfnamefont{G.}~\bibnamefont{Bertone}},
  \bibinfo{author}{\bibfnamefont{D.}~\bibnamefont{Hooper}}, \bibnamefont{and}
  \bibinfo{author}{\bibfnamefont{J.}~\bibnamefont{Silk}},
  \bibinfo{journal}{Phys. Rept.} \textbf{\bibinfo{volume}{405}},
  \bibinfo{pages}{279} (\bibinfo{year}{2005}), \eprint{hep-ph/0404175}.

\bibitem[{\citenamefont{Hooper and Profumo}(2007)}]{Hooper:2007qk}
\bibinfo{author}{\bibfnamefont{D.}~\bibnamefont{Hooper}} \bibnamefont{and}
  \bibinfo{author}{\bibfnamefont{S.}~\bibnamefont{Profumo}},
  \bibinfo{journal}{Phys. Rept.} \textbf{\bibinfo{volume}{453}},
  \bibinfo{pages}{29} (\bibinfo{year}{2007}), \eprint{hep-ph/0701197}.

\bibitem[{\citenamefont{Feng}(2010)}]{Feng:2010gw}
\bibinfo{author}{\bibfnamefont{J.~L.} \bibnamefont{Feng}},
  \bibinfo{journal}{Ann. Rev. Astron. Astrophys.}
  \textbf{\bibinfo{volume}{48}}, \bibinfo{pages}{495} (\bibinfo{year}{2010}),
  \eprint{1003.0904}.

\bibitem[{\citenamefont{Chang et~al.}(2008)}]{Chang:2008aa}
\bibinfo{author}{\bibfnamefont{J.}~\bibnamefont{Chang}} \bibnamefont{et~al.},
  \bibinfo{journal}{Nature} \textbf{\bibinfo{volume}{456}},
  \bibinfo{pages}{362} (\bibinfo{year}{2008}).

\bibitem[{\citenamefont{Adriani et~al.}(2009)}]{Adriani:2008zr}
\bibinfo{author}{\bibfnamefont{O.}~\bibnamefont{Adriani}} \bibnamefont{et~al.}
  (\bibinfo{collaboration}{PAMELA}), \bibinfo{journal}{Nature}
  \textbf{\bibinfo{volume}{458}}, \bibinfo{pages}{607} (\bibinfo{year}{2009}),
  \eprint{0810.4995}.

\bibitem[{\citenamefont{Adriani et~al.}(2011)}]{Adriani:2011xv}
\bibinfo{author}{\bibfnamefont{O.}~\bibnamefont{Adriani}} \bibnamefont{et~al.}
  (\bibinfo{collaboration}{PAMELA}), \bibinfo{journal}{Phys. Rev. Lett.}
  \textbf{\bibinfo{volume}{106}}, \bibinfo{pages}{201101}
  (\bibinfo{year}{2011}), \eprint{1103.2880}.

\bibitem[{\citenamefont{Ackermann et~al.}(2012)}]{FermiLAT:2011ab}
\bibinfo{author}{\bibfnamefont{M.}~\bibnamefont{Ackermann}}
  \bibnamefont{et~al.} (\bibinfo{collaboration}{Fermi-LAT}),
  \bibinfo{journal}{Phys. Rev. Lett.} \textbf{\bibinfo{volume}{108}},
  \bibinfo{pages}{011103} (\bibinfo{year}{2012}), \eprint{1109.0521}.

\bibitem[{\citenamefont{Aguilar et~al.}(2013)}]{Aguilar:2013qda}
\bibinfo{author}{\bibfnamefont{M.}~\bibnamefont{Aguilar}} \bibnamefont{et~al.}
  (\bibinfo{collaboration}{AMS}), \bibinfo{journal}{Phys. Rev. Lett.}
  \textbf{\bibinfo{volume}{110}}, \bibinfo{pages}{141102}
  (\bibinfo{year}{2013}).

\bibitem[{\citenamefont{Aguilar et~al.}(2014)}]{Aguilar:2014fea}
\bibinfo{author}{\bibfnamefont{M.}~\bibnamefont{Aguilar}} \bibnamefont{et~al.}
  (\bibinfo{collaboration}{AMS}), \bibinfo{journal}{Phys. Rev. Lett.}
  \textbf{\bibinfo{volume}{113}}, \bibinfo{pages}{221102}
  (\bibinfo{year}{2014}).

\bibitem[{\citenamefont{Goodenough and Hooper}(2009)}]{Goodenough:2009gk}
\bibinfo{author}{\bibfnamefont{L.}~\bibnamefont{Goodenough}} \bibnamefont{and}
  \bibinfo{author}{\bibfnamefont{D.}~\bibnamefont{Hooper}}
  (\bibinfo{year}{2009}), \eprint{0910.2998}.

\bibitem[{\citenamefont{{Vitale} et~al.}(2009)\citenamefont{{Vitale},
  {Morselli}, and {for the Fermi/LAT Collaboration}}}]{2009arXiv0912.3828V}
\bibinfo{author}{\bibfnamefont{V.}~\bibnamefont{{Vitale}}},
  \bibinfo{author}{\bibfnamefont{A.}~\bibnamefont{{Morselli}}},
  \bibnamefont{and} \bibinfo{author}{\bibnamefont{{for the Fermi/LAT
  Collaboration}}}, \bibinfo{journal}{ArXiv e-prints}  (\bibinfo{year}{2009}),
  \eprint{0912.3828}.

\bibitem[{\citenamefont{Hooper and Goodenough}(2011)}]{Hooper:2010mq}
\bibinfo{author}{\bibfnamefont{D.}~\bibnamefont{Hooper}} \bibnamefont{and}
  \bibinfo{author}{\bibfnamefont{L.}~\bibnamefont{Goodenough}},
  \bibinfo{journal}{Phys. Lett.} \textbf{\bibinfo{volume}{B697}},
  \bibinfo{pages}{412} (\bibinfo{year}{2011}), \eprint{1010.2752}.

\bibitem[{\citenamefont{Hooper and Linden}(2011)}]{Hooper:2010im}
\bibinfo{author}{\bibfnamefont{D.}~\bibnamefont{Hooper}} \bibnamefont{and}
  \bibinfo{author}{\bibfnamefont{T.}~\bibnamefont{Linden}},
  \bibinfo{journal}{Phys. Rev.} \textbf{\bibinfo{volume}{D83}},
  \bibinfo{pages}{083517} (\bibinfo{year}{2011}), \eprint{1011.4520}.

\bibitem[{\citenamefont{Abazajian and Kaplinghat}(2012)}]{Abazajian:2012pn}
\bibinfo{author}{\bibfnamefont{K.~N.} \bibnamefont{Abazajian}}
  \bibnamefont{and}
  \bibinfo{author}{\bibfnamefont{M.}~\bibnamefont{Kaplinghat}},
  \bibinfo{journal}{Phys. Rev.} \textbf{\bibinfo{volume}{D86}},
  \bibinfo{pages}{083511} (\bibinfo{year}{2012}), \bibinfo{note}{[Erratum:
  Phys. Rev.D87,129902(2013)]}, \eprint{1207.6047}.

\bibitem[{\citenamefont{Gordon and Macias}(2013)}]{Gordon:2013vta}
\bibinfo{author}{\bibfnamefont{C.}~\bibnamefont{Gordon}} \bibnamefont{and}
  \bibinfo{author}{\bibfnamefont{O.}~\bibnamefont{Macias}},
  \bibinfo{journal}{Phys. Rev.} \textbf{\bibinfo{volume}{D88}},
  \bibinfo{pages}{083521} (\bibinfo{year}{2013}), \bibinfo{note}{[Erratum:
  Phys. Rev.D89,no.4,049901(2014)]}, \eprint{1306.5725}.

\bibitem[{\citenamefont{Hooper and Slatyer}(2013)}]{Hooper:2013rwa}
\bibinfo{author}{\bibfnamefont{D.}~\bibnamefont{Hooper}} \bibnamefont{and}
  \bibinfo{author}{\bibfnamefont{T.~R.} \bibnamefont{Slatyer}},
  \bibinfo{journal}{Phys. Dark Univ.} \textbf{\bibinfo{volume}{2}},
  \bibinfo{pages}{118} (\bibinfo{year}{2013}), \eprint{1302.6589}.

\bibitem[{\citenamefont{Daylan et~al.}(2014)\citenamefont{Daylan, Finkbeiner,
  Hooper, Linden, Portillo, Rodd, and Slatyer}}]{Daylan:2014rsa}
\bibinfo{author}{\bibfnamefont{T.}~\bibnamefont{Daylan}},
  \bibinfo{author}{\bibfnamefont{D.~P.} \bibnamefont{Finkbeiner}},
  \bibinfo{author}{\bibfnamefont{D.}~\bibnamefont{Hooper}},
  \bibinfo{author}{\bibfnamefont{T.}~\bibnamefont{Linden}},
  \bibinfo{author}{\bibfnamefont{S.~K.~N.} \bibnamefont{Portillo}},
  \bibinfo{author}{\bibfnamefont{N.~L.} \bibnamefont{Rodd}}, \bibnamefont{and}
  \bibinfo{author}{\bibfnamefont{T.~R.} \bibnamefont{Slatyer}}
  (\bibinfo{year}{2014}), \eprint{1402.6703}.

\bibitem[{\citenamefont{Zhou et~al.}(2015)\citenamefont{Zhou, Liang, Huang, Li,
  Fan, Feng, and Chang}}]{Zhou:2014lva}
\bibinfo{author}{\bibfnamefont{B.}~\bibnamefont{Zhou}},
  \bibinfo{author}{\bibfnamefont{Y.-F.} \bibnamefont{Liang}},
  \bibinfo{author}{\bibfnamefont{X.}~\bibnamefont{Huang}},
  \bibinfo{author}{\bibfnamefont{X.}~\bibnamefont{Li}},
  \bibinfo{author}{\bibfnamefont{Y.-Z.} \bibnamefont{Fan}},
  \bibinfo{author}{\bibfnamefont{L.}~\bibnamefont{Feng}}, \bibnamefont{and}
  \bibinfo{author}{\bibfnamefont{J.}~\bibnamefont{Chang}},
  \bibinfo{journal}{Phys. Rev.} \textbf{\bibinfo{volume}{D91}},
  \bibinfo{pages}{123010} (\bibinfo{year}{2015}), \eprint{1406.6948}.

\bibitem[{\citenamefont{Calore et~al.}(2015)\citenamefont{Calore, Cholis, and
  Weniger}}]{Calore:2014xka}
\bibinfo{author}{\bibfnamefont{F.}~\bibnamefont{Calore}},
  \bibinfo{author}{\bibfnamefont{I.}~\bibnamefont{Cholis}}, \bibnamefont{and}
  \bibinfo{author}{\bibfnamefont{C.}~\bibnamefont{Weniger}},
  \bibinfo{journal}{JCAP} \textbf{\bibinfo{volume}{1503}}, \bibinfo{pages}{038}
  (\bibinfo{year}{2015}), \eprint{1409.0042}.

\bibitem[{\citenamefont{Huang et~al.}(2015)\citenamefont{Huang, Enßlin, and
  Selig}}]{Huang:2015rlu}
\bibinfo{author}{\bibfnamefont{X.}~\bibnamefont{Huang}},
  \bibinfo{author}{\bibfnamefont{T.}~\bibnamefont{Enßlin}}, \bibnamefont{and}
  \bibinfo{author}{\bibfnamefont{M.}~\bibnamefont{Selig}}
  (\bibinfo{year}{2015}), \eprint{1511.02621}.

\bibitem[{\citenamefont{Ajello et~al.}(2015)}]{TheFermi-LAT:2015kwa}
\bibinfo{author}{\bibfnamefont{M.}~\bibnamefont{Ajello}} \bibnamefont{et~al.}
  (\bibinfo{collaboration}{Fermi-LAT}) (\bibinfo{year}{2015}),
  \eprint{1511.02938}.

\bibitem[{\citenamefont{Lake}(1990)}]{Lake:1990du}
\bibinfo{author}{\bibfnamefont{G.}~\bibnamefont{Lake}},
  \bibinfo{journal}{Nature} \textbf{\bibinfo{volume}{346}}, \bibinfo{pages}{39}
  (\bibinfo{year}{1990}).

\bibitem[{\citenamefont{Baltz and Wai}(2004)}]{Baltz:2004bb}
\bibinfo{author}{\bibfnamefont{E.~A.} \bibnamefont{Baltz}} \bibnamefont{and}
  \bibinfo{author}{\bibfnamefont{L.}~\bibnamefont{Wai}},
  \bibinfo{journal}{Phys. Rev.} \textbf{\bibinfo{volume}{D70}},
  \bibinfo{pages}{023512} (\bibinfo{year}{2004}), \eprint{astro-ph/0403528}.

\bibitem[{\citenamefont{Strigari}(2013)}]{Strigari:2013iaa}
\bibinfo{author}{\bibfnamefont{L.~E.} \bibnamefont{Strigari}},
  \bibinfo{journal}{Phys. Rept.} \textbf{\bibinfo{volume}{531}},
  \bibinfo{pages}{1} (\bibinfo{year}{2013}), \eprint{1211.7090}.

\bibitem[{\citenamefont{Simon and Geha}(2007)}]{Simon:2007dq}
\bibinfo{author}{\bibfnamefont{J.~D.} \bibnamefont{Simon}} \bibnamefont{and}
  \bibinfo{author}{\bibfnamefont{M.}~\bibnamefont{Geha}},
  \bibinfo{journal}{Astrophys. J.} \textbf{\bibinfo{volume}{670}},
  \bibinfo{pages}{313} (\bibinfo{year}{2007}), \eprint{0706.0516}.

\bibitem[{\citenamefont{York et~al.}(2000)}]{York:2000gk}
\bibinfo{author}{\bibfnamefont{D.~G.} \bibnamefont{York}} \bibnamefont{et~al.}
  (\bibinfo{collaboration}{SDSS}), \bibinfo{journal}{Astron. J.}
  \textbf{\bibinfo{volume}{120}}, \bibinfo{pages}{1579} (\bibinfo{year}{2000}),
  \eprint{astro-ph/0006396}.

\bibitem[{\citenamefont{McConnachie}(2012)}]{McConnachie:2012vd}
\bibinfo{author}{\bibfnamefont{A.~W.} \bibnamefont{McConnachie}},
  \bibinfo{journal}{Astron. J.} \textbf{\bibinfo{volume}{144}},
  \bibinfo{pages}{4} (\bibinfo{year}{2012}), \eprint{1204.1562}.

\bibitem[{\citenamefont{Ackermann et~al.}(2011)}]{Ackermann:2011wa}
\bibinfo{author}{\bibfnamefont{M.}~\bibnamefont{Ackermann}}
  \bibnamefont{et~al.} (\bibinfo{collaboration}{Fermi-LAT}),
  \bibinfo{journal}{Phys. Rev. Lett.} \textbf{\bibinfo{volume}{107}},
  \bibinfo{pages}{241302} (\bibinfo{year}{2011}), \eprint{1108.3546}.

\bibitem[{\citenamefont{Geringer-Sameth and
  Koushiappas}(2011)}]{GeringerSameth:2011iw}
\bibinfo{author}{\bibfnamefont{A.}~\bibnamefont{Geringer-Sameth}}
  \bibnamefont{and} \bibinfo{author}{\bibfnamefont{S.~M.}
  \bibnamefont{Koushiappas}}, \bibinfo{journal}{Phys. Rev. Lett.}
  \textbf{\bibinfo{volume}{107}}, \bibinfo{pages}{241303}
  (\bibinfo{year}{2011}), \eprint{1108.2914}.

\bibitem[{\citenamefont{Tsai et~al.}(2013)\citenamefont{Tsai, Yuan, and
  Huang}}]{Tsai:2012cs}
\bibinfo{author}{\bibfnamefont{Y.-L.~S.} \bibnamefont{Tsai}},
  \bibinfo{author}{\bibfnamefont{Q.}~\bibnamefont{Yuan}}, \bibnamefont{and}
  \bibinfo{author}{\bibfnamefont{X.}~\bibnamefont{Huang}},
  \bibinfo{journal}{JCAP} \textbf{\bibinfo{volume}{1303}}, \bibinfo{pages}{018}
  (\bibinfo{year}{2013}), \eprint{1212.3990}.

\bibitem[{\citenamefont{Mazziotta et~al.}(2012)\citenamefont{Mazziotta,
  Loparco, de~Palma, and Giglietto}}]{Mazziotta:2012ux}
\bibinfo{author}{\bibfnamefont{M.~N.} \bibnamefont{Mazziotta}},
  \bibinfo{author}{\bibfnamefont{F.}~\bibnamefont{Loparco}},
  \bibinfo{author}{\bibfnamefont{F.}~\bibnamefont{de~Palma}}, \bibnamefont{and}
  \bibinfo{author}{\bibfnamefont{N.}~\bibnamefont{Giglietto}},
  \bibinfo{journal}{Astropart. Phys.} \textbf{\bibinfo{volume}{37}},
  \bibinfo{pages}{26} (\bibinfo{year}{2012}), \eprint{1203.6731}.

\bibitem[{\citenamefont{{Cholis} and {Salucci}}(2012)}]{2012PhRvD..86b3528C}
\bibinfo{author}{\bibfnamefont{I.}~\bibnamefont{{Cholis}}} \bibnamefont{and}
  \bibinfo{author}{\bibfnamefont{P.}~\bibnamefont{{Salucci}}},
  \bibinfo{journal}{\prd} \textbf{\bibinfo{volume}{86}}, \bibinfo{eid}{023528}
  (\bibinfo{year}{2012}), \eprint{1203.2954}.

\bibitem[{\citenamefont{Ackermann et~al.}(2014)}]{Ackermann:2013yva}
\bibinfo{author}{\bibfnamefont{M.}~\bibnamefont{Ackermann}}
  \bibnamefont{et~al.} (\bibinfo{collaboration}{Fermi-LAT}),
  \bibinfo{journal}{Phys. Rev.} \textbf{\bibinfo{volume}{D89}},
  \bibinfo{pages}{042001} (\bibinfo{year}{2014}), \eprint{1310.0828}.

\bibitem[{\citenamefont{Ackermann
  et~al.}(2015{\natexlab{a}})}]{Ackermann:2015zua}
\bibinfo{author}{\bibfnamefont{M.}~\bibnamefont{Ackermann}}
  \bibnamefont{et~al.} (\bibinfo{collaboration}{Fermi-LAT})
  (\bibinfo{year}{2015}{\natexlab{a}}), \eprint{1503.02641}.

\bibitem[{\citenamefont{{Geringer-Sameth}
  et~al.}(2015)\citenamefont{{Geringer-Sameth}, {Koushiappas}, and
  {Walker}}}]{2015PhRvD..91h3535G}
\bibinfo{author}{\bibfnamefont{A.}~\bibnamefont{{Geringer-Sameth}}},
  \bibinfo{author}{\bibfnamefont{S.~M.} \bibnamefont{{Koushiappas}}},
  \bibnamefont{and} \bibinfo{author}{\bibfnamefont{M.~G.}
  \bibnamefont{{Walker}}}, \bibinfo{journal}{\prd}
  \textbf{\bibinfo{volume}{91}}, \bibinfo{eid}{083535} (\bibinfo{year}{2015}),
  \eprint{1410.2242}.

\bibitem[{\citenamefont{{Baring} et~al.}(2015)\citenamefont{{Baring}, {Ghosh},
  {Queiroz}, and {Sinha}}}]{2015arXiv151000389B}
\bibinfo{author}{\bibfnamefont{M.~G.} \bibnamefont{{Baring}}},
  \bibinfo{author}{\bibfnamefont{T.}~\bibnamefont{{Ghosh}}},
  \bibinfo{author}{\bibfnamefont{F.~S.} \bibnamefont{{Queiroz}}},
  \bibnamefont{and} \bibinfo{author}{\bibfnamefont{K.}~\bibnamefont{{Sinha}}},
  \bibinfo{journal}{ArXiv e-prints}  (\bibinfo{year}{2015}),
  \eprint{1510.00389}.

\bibitem[{\citenamefont{Abbott et~al.}(2005)}]{Abbott:2005bi}
\bibinfo{author}{\bibfnamefont{T.}~\bibnamefont{Abbott}} \bibnamefont{et~al.}
  (\bibinfo{collaboration}{Dark Energy Survey}) (\bibinfo{year}{2005}),
  \eprint{astro-ph/0510346}.

\bibitem[{\citenamefont{{Dark Energy Survey Collaboration}
  et~al.}(2016)\citenamefont{{Dark Energy Survey Collaboration}, {Abbott},
  {Abdalla}, {Allam}, {Aleksic}, {Amara}, {Bacon}, {Balbinot}, {Banerji},
  {Bechtol} et~al.}}]{2016arXiv160100329D}
\bibinfo{author}{\bibnamefont{{Dark Energy Survey Collaboration}}},
  \bibinfo{author}{\bibfnamefont{T.}~\bibnamefont{{Abbott}}},
  \bibinfo{author}{\bibfnamefont{F.~B.} \bibnamefont{{Abdalla}}},
  \bibinfo{author}{\bibfnamefont{S.}~\bibnamefont{{Allam}}},
  \bibinfo{author}{\bibfnamefont{J.}~\bibnamefont{{Aleksic}}},
  \bibinfo{author}{\bibfnamefont{A.}~\bibnamefont{{Amara}}},
  \bibinfo{author}{\bibfnamefont{D.}~\bibnamefont{{Bacon}}},
  \bibinfo{author}{\bibfnamefont{E.}~\bibnamefont{{Balbinot}}},
  \bibinfo{author}{\bibfnamefont{M.}~\bibnamefont{{Banerji}}},
  \bibinfo{author}{\bibfnamefont{K.}~\bibnamefont{{Bechtol}}},
  \bibnamefont{et~al.}, \bibinfo{journal}{ArXiv e-prints}
  (\bibinfo{year}{2016}), \eprint{1601.00329}.

\bibitem[{\citenamefont{Bechtol et~al.}(2015)}]{Bechtol:2015cbp}
\bibinfo{author}{\bibfnamefont{K.}~\bibnamefont{Bechtol}} \bibnamefont{et~al.}
  (\bibinfo{collaboration}{DES}), \bibinfo{journal}{Astrophys. J.}
  \textbf{\bibinfo{volume}{807}}, \bibinfo{pages}{50} (\bibinfo{year}{2015}),
  \eprint{1503.02584}.

\bibitem[{\citenamefont{Koposov et~al.}(2015)\citenamefont{Koposov, Belokurov,
  Torrealba, and Evans}}]{Koposov:2015cua}
\bibinfo{author}{\bibfnamefont{S.~E.} \bibnamefont{Koposov}},
  \bibinfo{author}{\bibfnamefont{V.}~\bibnamefont{Belokurov}},
  \bibinfo{author}{\bibfnamefont{G.}~\bibnamefont{Torrealba}},
  \bibnamefont{and} \bibinfo{author}{\bibfnamefont{N.~W.} \bibnamefont{Evans}},
  \bibinfo{journal}{Astrophys. J.} \textbf{\bibinfo{volume}{805}},
  \bibinfo{pages}{130} (\bibinfo{year}{2015}), \eprint{1503.02079}.

\bibitem[{\citenamefont{Laevens et~al.}(2015)}]{Laevens:2015una}
\bibinfo{author}{\bibfnamefont{B.~P.~M.} \bibnamefont{Laevens}}
  \bibnamefont{et~al.}, \bibinfo{journal}{Astrophys. J.}
  \textbf{\bibinfo{volume}{802}}, \bibinfo{pages}{L18} (\bibinfo{year}{2015}),
  \eprint{1503.05554}.

\bibitem[{\citenamefont{{Martin} et~al.}(2015)\citenamefont{{Martin},
  {Nidever}, {Besla}, {Olsen}, {Walker}, {Vivas}, {Gruendl}, {Kaleida},
  {Mu{\~n}oz}, {Blum} et~al.}}]{martin15_hydra}
\bibinfo{author}{\bibfnamefont{N.~F.} \bibnamefont{{Martin}}},
  \bibinfo{author}{\bibfnamefont{D.~L.} \bibnamefont{{Nidever}}},
  \bibinfo{author}{\bibfnamefont{G.}~\bibnamefont{{Besla}}},
  \bibinfo{author}{\bibfnamefont{K.}~\bibnamefont{{Olsen}}},
  \bibinfo{author}{\bibfnamefont{A.~R.} \bibnamefont{{Walker}}},
  \bibinfo{author}{\bibfnamefont{A.~K.} \bibnamefont{{Vivas}}},
  \bibinfo{author}{\bibfnamefont{R.~A.} \bibnamefont{{Gruendl}}},
  \bibinfo{author}{\bibfnamefont{C.~C.} \bibnamefont{{Kaleida}}},
  \bibinfo{author}{\bibfnamefont{R.~R.} \bibnamefont{{Mu{\~n}oz}}},
  \bibinfo{author}{\bibfnamefont{R.~D.} \bibnamefont{{Blum}}},
  \bibnamefont{et~al.}, \bibinfo{journal}{\apjl}
  \textbf{\bibinfo{volume}{804}}, \bibinfo{eid}{L5} (\bibinfo{year}{2015}),
  \eprint{1503.06216}.

\bibitem[{\citenamefont{{Kim} et~al.}(2015)\citenamefont{{Kim}, {Jerjen},
  {Mackey}, {Da Costa}, and {Milone}}}]{kim15_pegasus}
\bibinfo{author}{\bibfnamefont{D.}~\bibnamefont{{Kim}}},
  \bibinfo{author}{\bibfnamefont{H.}~\bibnamefont{{Jerjen}}},
  \bibinfo{author}{\bibfnamefont{D.}~\bibnamefont{{Mackey}}},
  \bibinfo{author}{\bibfnamefont{G.~S.} \bibnamefont{{Da Costa}}},
  \bibnamefont{and} \bibinfo{author}{\bibfnamefont{A.~P.}
  \bibnamefont{{Milone}}}, \bibinfo{journal}{\apjl}
  \textbf{\bibinfo{volume}{804}}, \bibinfo{eid}{L44} (\bibinfo{year}{2015}),
  \eprint{1503.08268}.

\bibitem[{\citenamefont{{Kim} and {Jerjen}}(2015)}]{kim15_horologium}
\bibinfo{author}{\bibfnamefont{D.}~\bibnamefont{{Kim}}} \bibnamefont{and}
  \bibinfo{author}{\bibfnamefont{H.}~\bibnamefont{{Jerjen}}},
  \bibinfo{journal}{\apjl} \textbf{\bibinfo{volume}{808}}, \bibinfo{eid}{L39}
  (\bibinfo{year}{2015}), \eprint{1505.04948}.

\bibitem[{\citenamefont{{Laevens} et~al.}(2015)\citenamefont{{Laevens},
  {Martin}, {Bernard}, {Schlafly}, {Sesar}, {Rix}, {Bell}, {Ferguson},
  {Slater}, {Sweeney} et~al.}}]{laevens15_3dsph}
\bibinfo{author}{\bibfnamefont{B.~P.~M.} \bibnamefont{{Laevens}}},
  \bibinfo{author}{\bibfnamefont{N.~F.} \bibnamefont{{Martin}}},
  \bibinfo{author}{\bibfnamefont{E.~J.} \bibnamefont{{Bernard}}},
  \bibinfo{author}{\bibfnamefont{E.~F.} \bibnamefont{{Schlafly}}},
  \bibinfo{author}{\bibfnamefont{B.}~\bibnamefont{{Sesar}}},
  \bibinfo{author}{\bibfnamefont{H.-W.} \bibnamefont{{Rix}}},
  \bibinfo{author}{\bibfnamefont{E.~F.} \bibnamefont{{Bell}}},
  \bibinfo{author}{\bibfnamefont{A.~M.~N.} \bibnamefont{{Ferguson}}},
  \bibinfo{author}{\bibfnamefont{C.~T.} \bibnamefont{{Slater}}},
  \bibinfo{author}{\bibfnamefont{W.~E.} \bibnamefont{{Sweeney}}},
  \bibnamefont{et~al.}, \bibinfo{journal}{\apj} \textbf{\bibinfo{volume}{813}},
  \bibinfo{eid}{44} (\bibinfo{year}{2015}), \eprint{1507.07564}.

\bibitem[{\citenamefont{{Kirby} et~al.}(2015)\citenamefont{{Kirby}, {Cohen},
  {Simon}, and {Guhathakurta}}}]{kirby15_tri2}
\bibinfo{author}{\bibfnamefont{E.~N.} \bibnamefont{{Kirby}}},
  \bibinfo{author}{\bibfnamefont{J.~G.} \bibnamefont{{Cohen}}},
  \bibinfo{author}{\bibfnamefont{J.~D.} \bibnamefont{{Simon}}},
  \bibnamefont{and}
  \bibinfo{author}{\bibfnamefont{P.}~\bibnamefont{{Guhathakurta}}},
  \bibinfo{journal}{\apjl} \textbf{\bibinfo{volume}{814}}, \bibinfo{eid}{L7}
  (\bibinfo{year}{2015}), \eprint{1510.03856}.

\bibitem[{\citenamefont{Geringer-Sameth
  et~al.}(2015)\citenamefont{Geringer-Sameth, Walker, Koushiappas, Koposov,
  Belokurov, Torrealba, and Evans}}]{Geringer-Sameth:2015lua}
\bibinfo{author}{\bibfnamefont{A.}~\bibnamefont{Geringer-Sameth}},
  \bibinfo{author}{\bibfnamefont{M.~G.} \bibnamefont{Walker}},
  \bibinfo{author}{\bibfnamefont{S.~M.} \bibnamefont{Koushiappas}},
  \bibinfo{author}{\bibfnamefont{S.~E.} \bibnamefont{Koposov}},
  \bibinfo{author}{\bibfnamefont{V.}~\bibnamefont{Belokurov}},
  \bibinfo{author}{\bibfnamefont{G.}~\bibnamefont{Torrealba}},
  \bibnamefont{and} \bibinfo{author}{\bibfnamefont{N.~W.} \bibnamefont{Evans}},
  \bibinfo{journal}{Phys. Rev. Lett.} \textbf{\bibinfo{volume}{115}},
  \bibinfo{pages}{081101} (\bibinfo{year}{2015}), \eprint{1503.02320}.

\bibitem[{\citenamefont{Hooper and Linden}(2015)}]{Hooper:2015ula}
\bibinfo{author}{\bibfnamefont{D.}~\bibnamefont{Hooper}} \bibnamefont{and}
  \bibinfo{author}{\bibfnamefont{T.}~\bibnamefont{Linden}},
  \bibinfo{journal}{JCAP} \textbf{\bibinfo{volume}{1509}}, \bibinfo{pages}{016}
  (\bibinfo{year}{2015}), \eprint{1503.06209}.

\bibitem[{\citenamefont{{Bonnivard} et~al.}(2015)\citenamefont{{Bonnivard},
  {Combet}, {Maurin}, {Geringer-Sameth}, {Koushiappas}, {Walker}, {Mateo},
  {Olszewski}, and {Bailey}}}]{2015ApJ...808L..36B}
\bibinfo{author}{\bibfnamefont{V.}~\bibnamefont{{Bonnivard}}},
  \bibinfo{author}{\bibfnamefont{C.}~\bibnamefont{{Combet}}},
  \bibinfo{author}{\bibfnamefont{D.}~\bibnamefont{{Maurin}}},
  \bibinfo{author}{\bibfnamefont{A.}~\bibnamefont{{Geringer-Sameth}}},
  \bibinfo{author}{\bibfnamefont{S.~M.} \bibnamefont{{Koushiappas}}},
  \bibinfo{author}{\bibfnamefont{M.~G.} \bibnamefont{{Walker}}},
  \bibinfo{author}{\bibfnamefont{M.}~\bibnamefont{{Mateo}}},
  \bibinfo{author}{\bibfnamefont{E.~W.} \bibnamefont{{Olszewski}}},
  \bibnamefont{and} \bibinfo{author}{\bibfnamefont{J.~I.}
  \bibnamefont{{Bailey}}, \bibfnamefont{III}}, \bibinfo{journal}{\apjl}
  \textbf{\bibinfo{volume}{808}}, \bibinfo{eid}{L36} (\bibinfo{year}{2015}),
  \eprint{1504.03309}.

\bibitem[{\citenamefont{Drlica-Wagner
  et~al.}(2015{\natexlab{a}})}]{Drlica-Wagner:2015xua}
\bibinfo{author}{\bibfnamefont{A.}~\bibnamefont{Drlica-Wagner}}
  \bibnamefont{et~al.} (\bibinfo{collaboration}{DES, Fermi-LAT}),
  \bibinfo{journal}{Astrophys. J.} \textbf{\bibinfo{volume}{809}},
  \bibinfo{pages}{L4} (\bibinfo{year}{2015}{\natexlab{a}}),
  \eprint{1503.02632}.

\bibitem[{\citenamefont{Drlica-Wagner
  et~al.}(2015{\natexlab{b}})}]{Drlica-Wagner:2015ufc}
\bibinfo{author}{\bibfnamefont{A.}~\bibnamefont{Drlica-Wagner}}
  \bibnamefont{et~al.} (\bibinfo{collaboration}{DES}),
  \bibinfo{journal}{Astrophys. J.} \textbf{\bibinfo{volume}{813}},
  \bibinfo{pages}{109} (\bibinfo{year}{2015}{\natexlab{b}}),
  \eprint{1508.03622}.

\bibitem[{\citenamefont{Atwood et~al.}(2013)}]{Atwood:2013rka}
\bibinfo{author}{\bibfnamefont{W.}~\bibnamefont{Atwood}} \bibnamefont{et~al.}
  (\bibinfo{collaboration}{Fermi-LAT}) (\bibinfo{year}{2013}),
  \eprint{1303.3514},
  \urlprefix\url{http://inspirehep.net/record/1223837/files/arXiv:1303.3514.pd%
f}.

\bibitem[{\citenamefont{Acero et~al.}(2015)}]{Acero:2015hja}
\bibinfo{author}{\bibfnamefont{F.}~\bibnamefont{Acero}} \bibnamefont{et~al.}
  (\bibinfo{collaboration}{Fermi-LAT}) (\bibinfo{year}{2015}),
  \eprint{1501.02003}.

\bibitem[{\citenamefont{Steigman et~al.}(2012)\citenamefont{Steigman, Dasgupta,
  and Beacom}}]{Steigman:2012nb}
\bibinfo{author}{\bibfnamefont{G.}~\bibnamefont{Steigman}},
  \bibinfo{author}{\bibfnamefont{B.}~\bibnamefont{Dasgupta}}, \bibnamefont{and}
  \bibinfo{author}{\bibfnamefont{J.~F.} \bibnamefont{Beacom}},
  \bibinfo{journal}{Phys. Rev.} \textbf{\bibinfo{volume}{D86}},
  \bibinfo{pages}{023506} (\bibinfo{year}{2012}), \eprint{1204.3622}.

\bibitem[{\citenamefont{{Wright} et~al.}(1994)\citenamefont{{Wright},
  {Griffith}, {Burke}, and {Ekers}}}]{1994ApJS...91..111W}
\bibinfo{author}{\bibfnamefont{A.~E.} \bibnamefont{{Wright}}},
  \bibinfo{author}{\bibfnamefont{M.~R.} \bibnamefont{{Griffith}}},
  \bibinfo{author}{\bibfnamefont{B.~F.} \bibnamefont{{Burke}}},
  \bibnamefont{and} \bibinfo{author}{\bibfnamefont{R.~D.}
  \bibnamefont{{Ekers}}}, \bibinfo{journal}{Astrophys. J. Suppl.}
  \textbf{\bibinfo{volume}{91}}, \bibinfo{pages}{111} (\bibinfo{year}{1994}).

\bibitem[{\citenamefont{Mauch et~al.}(2003)\citenamefont{Mauch, Murphy,
  Buttery, Curran, Hunstead, Piestrzynski, Robertson, and
  Sadler}}]{Mauch:2003zh}
\bibinfo{author}{\bibfnamefont{T.}~\bibnamefont{Mauch}},
  \bibinfo{author}{\bibfnamefont{T.}~\bibnamefont{Murphy}},
  \bibinfo{author}{\bibfnamefont{H.~J.} \bibnamefont{Buttery}},
  \bibinfo{author}{\bibfnamefont{J.}~\bibnamefont{Curran}},
  \bibinfo{author}{\bibfnamefont{R.~W.} \bibnamefont{Hunstead}},
  \bibinfo{author}{\bibfnamefont{B.}~\bibnamefont{Piestrzynski}},
  \bibinfo{author}{\bibfnamefont{J.~G.} \bibnamefont{Robertson}},
  \bibnamefont{and} \bibinfo{author}{\bibfnamefont{E.~M.}
  \bibnamefont{Sadler}}, \bibinfo{journal}{Mon. Not. Roy. Astron. Soc.}
  \textbf{\bibinfo{volume}{342}}, \bibinfo{pages}{1117} (\bibinfo{year}{2003}),
  \eprint{astro-ph/0303188}.

\bibitem[{\citenamefont{Ackermann
  et~al.}(2015{\natexlab{b}})}]{Ackermann:2015yfk}
\bibinfo{author}{\bibfnamefont{M.}~\bibnamefont{Ackermann}}
  \bibnamefont{et~al.} (\bibinfo{collaboration}{Fermi-LAT}),
  \bibinfo{journal}{Astrophys. J.} \textbf{\bibinfo{volume}{810}},
  \bibinfo{pages}{14} (\bibinfo{year}{2015}{\natexlab{b}}),
  \eprint{1501.06054}.

\bibitem[{\citenamefont{Liao et~al.}(2015)\citenamefont{Liao, Liang, Weng, Gu,
  and Fan}}]{Liao:2015jfj}
\bibinfo{author}{\bibfnamefont{N.-H.} \bibnamefont{Liao}},
  \bibinfo{author}{\bibfnamefont{Y.-F.} \bibnamefont{Liang}},
  \bibinfo{author}{\bibfnamefont{S.-S.} \bibnamefont{Weng}},
  \bibinfo{author}{\bibfnamefont{M.-F.} \bibnamefont{Gu}}, \bibnamefont{and}
  \bibinfo{author}{\bibfnamefont{Y.-Z.} \bibnamefont{Fan}}
  (\bibinfo{year}{2015}), \eprint{1510.05584}.

\bibitem[{\citenamefont{Cirelli}(2015)}]{Cirelli:2015gux}
\bibinfo{author}{\bibfnamefont{M.}~\bibnamefont{Cirelli}}
  (\bibinfo{year}{2015}), \eprint{1511.02031}.

\end{thebibliography}

\end{document}